\documentclass[pra,preprint,floatfix]{revtex4}
\usepackage{amssymb,amsmath}
\usepackage{graphicx}
\begin{document}
\newcommand{\xbar}{\ensuremath{\bar{x}}}
\newcommand{\ybar}{\ensuremath{\bar{y}}}
\newcommand{\xp}{\ensuremath{x^+}}
\newcommand{\xm}{\ensuremath{x^-}}
\newcommand{\yp}{\ensuremath{y^+}}
\newcommand{\ym}{\ensuremath{y^-}}
\newcommand{\xo}{\ensuremath{x_0}}
\newcommand{\yo}{\ensuremath{y_0}}
\newcommand{\xobar}{\ensuremath{\bar{x_0}}}
\newcommand{\yobar}{\ensuremath{\bar{y_0}}}
\newcommand{\xpo}{\ensuremath{x_0^+}}
\newcommand{\xmo}{\ensuremath{x_0^-}}
\newcommand{\ypo}{\ensuremath{y_0^+}}
\newcommand{\ymo}{\ensuremath{y_0^-}}
\newcommand{\Xbar}{\ensuremath{\bar{X}}}
\newcommand{\Ybar}{\ensuremath{\bar{Y}}}
\newcommand{\Xpls}{\ensuremath{X^+}}
\newcommand{\Xmns}{\ensuremath{X^-}}
\newcommand{\Ypls}{\ensuremath{Y^+}}
\newcommand{\Ymns}{\ensuremath{Y^-}}
\newcommand{\Xo}{\ensuremath{X_0}}
\newcommand{\Yo}{\ensuremath{Y_0}}
\newcommand{\Xobar}{\ensuremath{\bar{X_0}}}
\newcommand{\Yobar}{\ensuremath{\bar{Y_0}}}
\newcommand{\Xopls}{\ensuremath{X_0^+}}
\newcommand{\Xomns}{\ensuremath{X_0^-}}
\newcommand{\Yopls}{\ensuremath{Y_0^+}}
\newcommand{\Yomns}{\ensuremath{Y_0^-}}
\newcommand{\Po}{\ensuremath{P_0}}
\newcommand{\Qo}{\ensuremath{Q_0}}
\newcommand{\po}{\ensuremath{p_0}}
\newcommand{\qo}{\ensuremath{q_0}}
\newcommand{\M}{\ensuremath{\mathcal{M}}}
\newcommand{\R}{\ensuremath{\mathcal{R}}}
\newcommand{\T}{\ensuremath{\mathcal{T}}}
\newcommand{\Sm}{\ensuremath{\mathcal{S}}}
\newcommand{\U}{\ensuremath{\mathcal{U}}}
\newcommand{\Q}{\ensuremath{\mathcal{Q}}}
\newcommand{\Qpm}{\ensuremath{\mathcal{Q}_{\pm}}}
\newcommand{\ket}[1]{\ensuremath{\left|#1\right\rangle}}
\newcommand{\bra}[1]{\ensuremath{\left\langle#1\right|}}
\newcommand{\braket}[2]{\ensuremath{\left\langle#1|#2\right\rangle}}
\newcommand{\expect}[1]{\ensuremath{\left\langle#1\right\rangle}}
\newcommand{\braopket}[3]{\ensuremath{\left\langle#1\right|#2\left|#3\right\rangle}}
\newcommand{\commutator}[2]{\ensuremath{\left[#1,#2\right]}}
\newcommand{\anticommutator}[2]{\ensuremath{\left\{#1,#2\right\}}}
\newcommand{\pdiff}[2]{\ensuremath{\frac{\partial #1}{\partial #2}}}
\newcommand{\diff}[2]{\ensuremath{\frac{d #1}{d #2}}}
\newcommand{\dpls}{\ensuremath{\partial_+}}
\newcommand{\dmns}{\ensuremath{\partial_-}}
\newcommand{\z}{\ensuremath{\vec{z}}}
\newcommand{\Z}{\ensuremath{\vec{Z}}}
\newcommand{\zpm}{\ensuremath{\vec{z}_{\pm}}}
\newcommand{\ct}{\ensuremath{\cos\theta}}
\newcommand{\cnt}[1]{\ensuremath{\cos^{#1}\theta}}
\newcommand{\st}{\ensuremath{\sin\theta}}
\newcommand{\snt}[1]{\ensuremath{\sin^{#1}\theta}}
\newcommand{\cwt}{\ensuremath{\cos(\omega t)}}
\newcommand{\cnwt}[1]{\ensuremath{\cos^{#1}(\omega t)}}
\newcommand{\swt}{\ensuremath{\sin(\omega t)}}
\newcommand{\snwt}[1]{\ensuremath{\sin^{#1}(\omega t)}}
\newcommand{\clt}{\ensuremath{\cos(\lambda t)}}
\newcommand{\cnlt}[1]{\ensuremath{\cos^{#1}(\lambda t)}}
\newcommand{\slt}{\ensuremath{\sin(\lambda t)}}
\newcommand{\snlt}[1]{\ensuremath{\sin^{#1}(\lambda t)}}
\newcommand{\chlt}{\ensuremath{\cosh(\lambda t)}}
\newcommand{\chnlt}[1]{\ensuremath{\cosh^{#1}(\lambda t)}}
\newcommand{\shlt}{\ensuremath{\sinh(\lambda t)}}
\newcommand{\shnlt}[1]{\ensuremath{\sinh^{#1}(\lambda t)}}
\newcommand{\cws}{\ensuremath{\cos(\omega s)}}
\newcommand{\sws}{\ensuremath{\sin(\omega s)}}
\newcommand{\cls}{\ensuremath{\cos(\lambda s)}}
\newcommand{\sls}{\ensuremath{\sin(\lambda s)}}
\newcommand{\chls}{\ensuremath{\cosh(\lambda s)}}
\newcommand{\shls}{\ensuremath{\sinh(\lambda s)}}
\newcommand{\cvec}[4]{\ensuremath{\left(\begin{array}{c}#1\\#2\\#3\\#4\end{array}\right)}}
\newcommand{\rvec}[4]{\ensuremath{\left(#1,#2,#3,#4\right)}}
\newcommand{\omeff}{\ensuremath{\omega_{\mbox{\tiny eff}}}}
\newcommand{\gameff}{\ensuremath{\gamma_{\mbox{\tiny eff}}}}
\newcommand{\Tr}{\ensuremath{\mbox{Tr}}}
\newcommand{\Ft}{\ensuremath{\hat{F}}}
\newcommand{\Denom}{\ensuremath{\left(\omega^2-\lambda^2\right)\cnt2\snt2\swt\shlt +
	\omega\lambda\left(2\cwt\chlt\cnt2\snt2+\cnt4+\snt4\right)}}

\newcommand{\syssymbol} {\ensuremath{\mathcal{S}}}
\newcommand{\envsymbol} {\ensuremath{\mathcal{E}}}
\newcommand{\HS} {\ensuremath{\hat{H}_{\syssymbol}}}
\newcommand{\HE} {\ensuremath{\hat{H}_{\envsymbol}}}
\newcommand{\HSE} {\ensuremath{\hat{H}_{\syssymbol\envsymbol}}}
\newcommand{\Hint} {\HSE}
\newcommand{\sysop} {\ensuremath{\hat{\mathcal{O}}_{\syssymbol}}}
\newcommand{\envop} {\ensuremath{\hat{\mathcal{O}}_{\envsymbol}}}
\newcommand{\sysstate}[1] {\ensuremath{\ket{\Psi_{\syssymbol}#1}}}
\newcommand{\envstate}[1] {\ensuremath{\ket{\Psi_{\envsymbol}#1}}}
\newcommand{\superstate}[1] {\ensuremath{\ket{\Psi_{\syssymbol\envsymbol}#1}}}
\newcommand{\syseigenstate}[2] {\ensuremath{\ket{s_{#1}#2}}}
\newcommand{\syseigenvalue} {\ensuremath{s}}
\newcommand{\enveigenvalue} {\ensuremath{\varepsilon}}
\newcommand{\enveigenstate}[2] {\ensuremath{\ket{\varepsilon_{#1}#2}}}
\newcommand{\varentropy} {\ensuremath{\tilde{\mathcal{S}}}}
\newcommand{\entropy} {S}
\newcommand{\energy} {E}
\newcommand{\enveigenstateoverlaps} {\ensuremath{\braket{\varepsilon_n(t)}{\varepsilon_m(t)}}}
\newcommand{\V}{\ensuremath{\mathcal{V}}}
\newcommand{\Vp}{\ensuremath{\mathcal{V}_p}}
\newcommand{\Ve}{\ensuremath{\mathcal{V}_\envsymbol}}

\title{Decoherence from a Chaotic Environment:  An Upside Down ``Oscillator'' as a Model}
\date{\today}
\author{Robin Blume-Kohout}
\email{rbk@socrates.berkeley.edu}
\author{Wojciech H. Zurek}
\email{whz@lanl.gov}
\affiliation{Los Alamos National Laboratory}
\begin{abstract}
Chaotic evolutions exhibit exponential sensitivity to initial conditions.  This suggests that even very small perturbations resulting from weak coupling of a quantum chaotic environment to the position of a system whose state is a non-local superposition will lead to rapid decoherence.  However, it is also known that quantum counterparts of classically chaotic systems lose exponential sensitivity to initial conditions, so this expectation of enhanced decoherence is by no means obvious.  We analyze decoherence due to a ``toy'' quantum environment that is analytically solvable, yet displays the crucial phenomenon of exponential sensitivity to perturbations.  We show that such an environment, with a single degree of freedom, can be far more effective at destroying quantum coherence than a heat bath with infinitely many degrees of freedom.  This also means that the standard ``quantum Brownian motion'' model for a decohering environment may not be as universally applicable as it once was conjectured to be.
\end{abstract}
\maketitle

\section{Introduction and Motivation}
Chaotic classical systems display a phenomenon known as \emph{sensitive dependence on initial conditions}.  Two copies of such a system, prepared in nearly identical states (two distinct points in phase space, separated by a very small distance), will evolve over time into widely separated states.  In an idealized case, the distance between the two points in phase space grows as $e^{\lambda t}$, where $\lambda$ is the largest \emph{Lyapunov exponent} of the system.  This does not happen in quantum mechanics.  The (oversimplified) reason is that quantum mechanics is linear; thus two ``nearly identical'' states (i.e., states with a large initial overlap) remain nearly identical -- their overlap is constant under unitary evolution -- for all time!  However, Peres \cite{Peres95Book} predicted an \emph{analogous} phenomenon for quantum systems: two nearly identical systems, prepared in identical states but obeying slightly different Hamiltonians (i.e., $\hat{H}$ and $\hat{H}+\delta\hat{H}$), will evolve into two different states whose inner product decays exponentially in time when $\hat{H}$ is the quantization of a chaotic Hamiltonian.  This idea has recently been studied and extended by Jalabert and Pastawski \cite{JalabertPastawski01PRL}.

Our interest in this behavior stems from a simple model of decoherence, the process by which most pure states evolve into mixtures due to interaction with an environment \cite{Zurek91PhysicsToday, Giulini96Book, PazZurek01LesHouches, Zurek01Nature, Zurek03RMP}.  A simple but general model of decoherence is generated by an overall Hamiltonian $\HS+\HE+\HSE$, where the interaction Hamiltonian $\HSE$ is a product of a system operator $\sysop$ and an environment operator $\envop$.  When the system is in an eigenstate $\syseigenstate{n}{}$ of $\sysop$ (with eigenvalue $\syseigenvalue_n$), then the environment evolves according to an effective Hamiltonian $\HE + \syseigenvalue_n\envop$.  More generally, the system will be in a superposition of the eigenstates of $\sysop$, in which case the environment experiences a Hamiltonian that is conditional on the state of the system.  Thus, if the initial (unentangled) state of the supersystem is $\superstate{(0)} = \left(\sum_n{\alpha_n\syseigenstate{n}{(0)}}\right)\envstate{(0)}$ (where $\envstate{(0)}$ is simply the initial state of the environment), then after a time $t$ the state will become $\superstate{(t)} = \sum_n{\alpha_n\syseigenstate{n}{(t)}\enveigenstate{n}{(t)}}$, where $\enveigenstate{n}{(t)}$ is the state into which the environment evolves \emph{if} the system is in state $\syseigenstate{n}{}$ (equation (\ref{EnvEvolution})).  When $|\enveigenstateoverlaps| = 1$, no entanglement occurs, the system remains in a pure state, and there is no decoherence.  When, however, $|\enveigenstateoverlaps| \sim 0$, superpositions of $\syseigenstate{n}{}$ and $\syseigenstate{m}{}$ are transformed into completely decohered mixtures.

This model is tremendously simplified -- it ignores many important phenomena, including the effect of the system Hamiltonian.  However, it illustrates a key point: the rate at which an environment induces decoherence in a system depends on the decay rate of the overlap $|\enveigenstateoverlaps|$ , where
\begin{equation}
\enveigenstate{j}{(t)} = \exp{\left(i(\HE + \syseigenvalue_n{}\envop)t\right)}\envstate{(0)}
\label{EnvEvolution}
\end{equation}
Thus, we expect that the sensitivity to perturbation of a chaotic environment's Hamiltonian will result in rapid decoherence even for very weak couplings.  This conclusion is encouraged by related analytic \cite{Zurek01Nature, KarkuszewskiJarzynskiZurek02PRL} and numerical \cite{SakagamiKubotaniOkamura96ProgTP, KubotaniOkamuraSakagami95PhysicaA, MillerSarkar99PRE} studies.  Some of its aspects are also beginning to be investigated experimentally \cite{JalabertPastawski01PRL, PastawskiLevsteinUsaj00PhysicaA, UsajPastawskiLevstein98MolPhys, LevsteinUsajPastawski98JCP}.  The nature of the chaotic evolution seems to be important for this conclusion \cite{GorinSeligman02JOpticsB, ProsenSeligman02JPhysA}, and the physics of the related phenomena is still being debated \cite{JacquodSilverstrovBeenakker02PRL, JacquodSilverstrovBeenakker01PRE}.  It is therefore useful to have an exactly solvable model that captures \emph{some} of the features of quantum chaotic evolutions.

A well-known feature of completely chaotic classical systems is that at every point in phase space there exist stable and unstable manifolds -- directions along which a cell respectively shrinks and grows exponentially.  These manifolds fold in phase space, enabling a given region to be always stretching in some direction, yet remain within a bounded volume.  We examine a system that exhibits such an exponential sensitivity to initial conditions, yet is analytically solvable: the inverted harmonic oscillator.  We note that our model does not exhibit folding; we shall comment on the consequences of this shortcoming in due course.  We shall also not discuss models with mixed phase spaces -- our model is clearly too simple-minded for that.

\section{Analysis of the Unstable Oscillator}

The Hamiltonian for an inverted oscillator is $H = \frac{p^2}{2M} - \frac{M\lambda^2x^2}{2}$, where we have replaced the parameter $\omega^2$ for a simple harmonic oscillator (SHO) with $-\lambda^2$.  Such an ``oscillator'' does not oscillate at all; it has an unstable fixed point at $\{x=0,p=0\}$, but accelerates exponentially away from the fixed point when perturbed.  The price we pay for a solvable system that displays such sensitivity is that its phase space is unbounded; the kinetic energy of the system grows approximately as $E \propto e^{2\lambda t}$.  Although such a system is clearly nonphysical, it is an excellent short-time approximation for some real unstable systems.  We believe that it can reproduce some of the behavior of a chaotic environment.  To be specific, our model can be thought of as approximating the exponential sensitivity exhibited by the chaotic systems.  In real chaotic systems, however, in addition to the folding we have already mentioned, local values of chaotic exponents can be quite different from their time-averages.  Moreover, the directions of stable and unstable manifolds can vary from point to point.  Again, our model misses this feature.  On the other hand, we are encouraged by the fact that this ``integrable model of chaos'' has been used previously \cite{ZurekPaz94PRL}, in a different context, and that its conclusions seem to be confirmed by numerical simulations \cite{KubotaniOkamuraSakagami95PhysicaA, SakagamiKubotaniOkamura96ProgTP, MillerSarkarZarum98APPB, HabibShizumeZurek98PRL, MillerSarkar99PRE, MonteolivaPaz00PRL, MonteolivaPaz01PRE, BianucciPazSaraceno02PRE}.

We consider an inverted harmonic environment (IHE) consisting of one such oscillator, and couple it to a system consisting of a single SHO with mass $m$ and frequency $\omega$.  This supersystem is linear, so it can be analyzed with the same master equation techniques used to treat quantum Brownian motion (QBM) and other linear problems.  We will briefly discuss the novel method we use to obtain the coefficients of the master equation.

\subsection{Obtaining the Coefficients of the Master Equation}
We begin with the standard master equation for a linear system coupled linearly to a linear environment (as derived in \cite{CaldeiraLeggett83PhysicaA, UnruhZurek89PRD,HuPazZhang92PRD}),
\begin{widetext}
\begin{equation}
\pdiff{}{t}\hat{\rho} = \begin{array}{l}
        \frac{1}{i\hbar}\left(\frac{m\omeff^2}{2}\commutator{\hat{x}^2}{\hat{\rho}}
        +\frac{1}{2m}\commutator{\hat{p}^2}{\hat{\rho}}
        +\frac{\gameff}{2}\commutator{\hat{x}}{\anticommutator{\hat{p}}{\hat{\rho}}}
        -F(t)\commutator{\hat{x}}{\hat{\rho}}\right)\\
        -f_1(t)\commutator{\hat{x}}{\commutator{\hat{x}}{\hat{\rho}}}
        +f_2(t)\commutator{\hat{x}}{\commutator{\hat{p}}{\hat{\rho}}}\end{array}.
	\label{MasterEquation}
\end{equation}
\end{widetext}
This form is exact when all the terms in $H=\HS+\HE+\HSE$ are linear and quadratic in positions and momenta, provided that the coupling terms in $\HSE$ involve only position operators.  Here and throughout, we use $x$ and $p$ to denote the position and momentum of the system, and $y_i$ and $q_i$ for those of the $i$th environmental degree of freedom.  To analyze a particular system-environment combination, we need to obtain the specific values of the time-dependent coefficients $\omeff^2,\gameff,F,f_1$, and $f_2$.  Although the IHE has a single degree of freedom, we consider an arbitrary linear environment as long as possible in order to make the generality of our method explicit.  At the end of the general analysis, we will specialize to the IHE.

Since the Heisenberg equations of motion for the system and environment operators match exactly the classical equations of motion for the equivalent variables, we can obtain the coefficients of the master equation from its classical analogue, the equation of motion for the reduced system.  This straightforward approach has been examined previously in \cite{AnglinHabib96MPLA}, but primarily in the context of QBM-like systems with infinitely many degrees of freedom.

The supersystem is linear and Hamiltonian, so we can write its trajectories as 
\begin{equation}
\z(t) = \T(t)\z(0) \label{ClassicalTrajectories},
\end{equation}
where $\z$ is a $2N$-dimensional vector of the form $[x,p,y_1,q_1,y_2,\ldots]$ and $\T(t)$ is a square $2N\times 2N$ matrix.  An equation of motion gives the derivatives of $\z(t)$ in terms of $\z(t)$ itself. We obtain this by first differentiating equation (\ref{ClassicalTrajectories}) to obtain $\dot{\z}(t) = \dot{\T}\z(0)$, then substituting the $\z(0)$ obtained by inverting equation (\ref{ClassicalTrajectories}): $\z(0) = \T^{-1}\z(t)$.  This yields
\begin{equation}
\dot{\z}(t) = \dot{\T}\T^{-1}\z(t) \label{ClassicalDerivatives}
\end{equation}
This is an equation of motion for the supersystem; it gives the time derivatives of all the supersystem coordinates and momenta in terms of their values at time $t$.  However, this is \emph{not} the equation of motion that we need.  When we trace over the environment to obtain the master equation, we assume that the environment's state at time $t$ is inaccessible; we know only its initial state.  We need a different equation, one that respects this constraint.

To obtain an equation of motion that provides $\dot{x}(t)$ and $\dot{p}(t)$ in terms of $x(t),p(t)$, and the initial state of the environment, we define a new matrix $\T_p$ related to $\T$:
\begin{equation}
(\T_p)_{ij} = \left\{\begin{array}{l}\T_{ij}\text{ for }i\in\{1,2\}\\
								  \delta_{ij}\text{ for }i>2\end{array}\right.
\end{equation}
As we evolve the quantum state of the system, we presume that at all times we have access to knowledge of (1) the reduced density matrix of the system, and (2) the \emph{initial} state of the environment.  The corresponding classical state of knowledge is a vector $\z_p(t) = [x(t),p(t),y_1(0),q_1(0),\ldots]$.  This vector can be obtained using $\T_p$: $\z_p(t) = \T_p(t)\z_p(0)$.  By the same process that led to equation (\ref{ClassicalDerivatives}), we conclude:
\begin{equation}
\dot{\z}_p(t) = \dot{\T_p}\T_p^{-1}\z_p(t). \label{ClassicalPDerivatives}
\end{equation}
This yields $\dot{x}(t)$ and $\dot{p}(t)$, but in order to obtain the coefficients in the master equation, we need the derivatives of higher order powers of $x$ and $p$.  This requires some simplifying assumption; throughout this paper, we will choose to assume that all states are Gaussian.  This is particularly convenient since Gaussian states form a closed set under linear evolution.  Such states are completely described by linear and quadratic expectation values of $x$ and $p$, so to characterize their evolution we need time derivatives of $x^2$, $p^2$, and $xp$ as well as $\dot{x}$ and $\dot{p}$.  This is straightforward; we define the symmetric variance tensor $\Vp=\z_p\z_p^T$, which contains all the quadratic combinations of $x$ and $p$, and transforms as $\Vp(t) = \T_p\Vp(0)\T_p^T$.  The time derivative of $\Vp$ is thus given by:
\begin{equation}
\dot{\Vp} = \dot{\T_p}\T_p^{-1}\Vp + \Vp\left(\dot{\T_p}\T_p^{-1}\right)^T.
\label{ClassicalVarianceDerivatives}
\end{equation}

Now, we need to relate these quantities to the coefficients of the master equation.  For any time-independent quantum operator $\hat{A}$, $\pdiff{}{t}\expect{\hat{A}} = \Tr{\left(\dot{\rho}\hat{A}\right)}$.  The derivatives of the relevant expectation values are obtained from the master equation:
\begin{widetext}
\begin{eqnarray}
\pdiff{\expect{x}}{t} &=& \frac{\expect{\hat{p}}}{m}\label{MEdxdt}\\
\pdiff{\expect{p}}{t} &=& -m\omeff^2\expect{\hat{x}}-\gameff\expect{\hat{p}}+F(t)\label{MEdpdt}\\
\pdiff{\expect{x^2}}{t} &=& \frac{2}{m}\expect{\{\hat{x},\hat{p}\}/2}\label{MEdxxdt}\\
\pdiff{\expect{p^2}}{t} &=& -2m\omega^2\expect{\{\hat{x},\hat{p}\}/2} - 2\gameff\expect{\hat{p}^2}
                        +2F(t)\expect{\hat{p}} + 2\hbar^2 f_1(t)\label{MEdppdt}\\
\pdiff{\expect{\{x,p\}/2}}{t} &=& -m\omeff^2\expect{\hat{x}^2} + \frac{1}{m}\expect{\hat{p}^2}
                        -\gameff\expect{\{\hat{x},\hat{p}\}/2} + F(t)\expect{\hat{x}} +\hbar^2 f_2(t)\label{MEdxpdt}
\end{eqnarray}
We can apply the preceding classical analysis to the Heisenberg operators, then equate the results with those from equations (\ref{MEdxdt}-\ref{MEdxpdt}).  Equations (\ref{MEdxdt}-\ref{MEdxpdt}) imply that the matrix $\dot{\T}_p\T_p^{-1}$ that gives the derivatives of the Heisenberg operators must take the form
\begin{equation}
\dot{\T}_p\T_p^{-1} = \left(\begin{array}{ccccc} 0 & 1/m & 0 & 0 & \ldots\\
	-m\omeff^2 & -\gameff & F_{y1} & F_{q1} & \ldots\\
	0 & 0 & 0 & 0 & \ldots\\
	0 & 0 & 0 & 0 & \ldots\\
	\vdots & \vdots & \vdots & \vdots &\ddots \end{array}\right),\label{MatrixForm}
\end{equation}
where the net force $F(t)$ is the expectation value of a force operator: $\Ft(t) = \sum_i{F_{yi}(t)\hat{y}_i(0) + F_{qi}(t)\hat{q}_i(0)}$.  Thus, $F(t) = \sum_i{F_{yi}\expect{\hat{y}_i} + F_{qi}\expect{\hat{q}_i}}$.  Using this matrix in equation (\ref{ClassicalVarianceDerivatives}) yields the time derivatives of the system variances $\expect{\hat{x}^2}$, $\expect{\hat{p}^2}$ , and $\expect{\{\hat{x},\hat{p}\}/2}$:
\begin{eqnarray}
\pdiff{\expect{\hat{x}^2}}{t} &=& \dot{\left(\Vp\right)}_{11} = \frac{2}{m}\expect{\{\hat{x},\hat{p}\}/2}\\
\pdiff{\expect{\hat{p}^2}}{t} &=& \dot{\left(\Vp\right)}_{22} = -2m\omeff^2\expect{\{\hat{x},\hat{p}\}/2} -
		2\gameff\expect{\hat{p}^2} + \expect{\{\Ft(t),\hat{p}\}}\\
\pdiff{\expect{\{\hat{x},\hat{p}\}/2}}{t} &=& 	
		\frac{\dot{\left(\Vp\right)}_{12}+\dot{\left(\Vp\right)}_{21}}{2}\nonumber\\
		&=& -m\omeff^2\expect{\hat{x}^2} + \frac{1}{m}\expect{\hat{p}^2} - \gameff\expect{\{\hat{x},\hat{p}\}/2} + 		
		\frac12\expect{\{\Ft(t),\hat{x}\}}.
\end{eqnarray}
\end{widetext}

Comparing this with the results from equations (\ref{MEdxdt}-\ref{MEdxpdt}), we can solve for $f_1$ and $f_2$ as
\begin{eqnarray}
f_1(t) &=& \frac{1}{2\hbar^2}\left(\expect{\{\Ft,\hat{p}(t)\}}-2\expect{\Ft}\expect{\hat{p}(t)}\right)\\
f_2(t) &=& \frac{1}{2\hbar^2}\left(\expect{\{\Ft,\hat{x}(t)\}}-2\expect{\Ft}\expect{\hat{x}(t)}\right)
\end{eqnarray}
Clearly, these coefficients are nonzero only if the force operator $\Ft$ is correlated with (respectively) $\hat{p}$ or $\hat{x}$.  Since $\Ft$ is always expressed in terms of the $t=0$ operators of the environment ($\hat{y}_i(0),\hat{q}_i(0)$), these correlations occur because the evolution mixes system and environment operators: $\hat{p}(t) = \sum_i{\T_{2i}\hat{z}_i(0)}$ and $\hat{x}(t) = \sum_i{\T_{1i}\hat{z}_i(0)}$.  By expanding $\hat{x}(t)$ and $\hat{p}(t)$ in this way and indulging in some tedious algebra, we obtain $f_1(t)$ and $f_2(t)$.  They are most conveniently expressed as the contraction of two tensors, one of which is the initial variance tensor of the environment,
\begin{equation}
\Ve = \left(\begin{array}{cc}\Delta y_i^2(0) & \Delta y_iq_i(0) \\ \Delta y_iq_i(0) & \Delta q_i^2(0)\end{array}\right). \label{EnvVarianceTensor}
\end{equation}
For lack of better notation, we use $\Delta y q \equiv \expect{\frac12\left(\hat{y}\hat{q}+\hat{q}\hat{y}\right)}-\expect{\hat{y}}\expect{\hat{q}}$ throughout.  Technically, this is a second cumulant, $\Delta yq = \left<\left<\hat{y}\hat{q}\right>\right>$, but we have adopted this notation because cumulant notation is not widely familiar. \label{CumulantNote}

In terms of $\Ve$, the diffusion coefficients are
\begin{widetext}
\begin{eqnarray}
\hbar^2 f_1 &=&\sum_{i=1}^N{\epsilon_i^{-1}\Tr\left[\left(\begin{array}{cc}m_iF_{yi}\T_{2,2i+1} & m_iF_{qi}\T_{2,2i+1}\\
	F_{y_i}\T_{2,2i+2} & F_{qi}\T_{2,2i+2}\end{array}\right)\Ve\right]}\label{Eqf1}\\
\hbar^2 f_2 &=&\sum_{i=1}^N{\epsilon_i^{-1}\Tr\left[\left(\begin{array}{cc}m_iF_{yi}\T_{1,2i+1} & m_iF_{qi}\T_{1,2i+1}\\
	F_{y_i}\T_{1,2i+2} & F_{qi}\T_{1,2i+2}\end{array}\right)\Ve\right]}\label{Eqf2}\\
\end{eqnarray}
\end{widetext}

We now have all the coefficients of the master equation in terms of elements of the two matrices $\T_p$ and $\dot{\T}_p\T_p^{-1}$ (note that while equations (\ref{Eqf1}-\ref{Eqf2}) are expressed in terms of $\T$ instead of $\T_p$, they involve only the first two rows, which are identical between $\T$ and $\T_p$).  While complete specification of $\T_p$ requires complete specification of the systems and their couplings, we can simplify the problem further.  The underlying physics mandates that the matrix $\T_p$ be of a particular form.  We first define the $2\times 2$ matrix
\begin{equation}
\M_i = \left(\begin{array}{cc}\sqrt{\frac{m_i}{m_s}}\dot{\phi}_i(t) & \frac{1}{\sqrt{m_i m_s}}\phi_i(t)\\ \\
	\sqrt{m_i m_s}\ddot{\phi}_i(t) & \sqrt{\frac{m_s}{m_i}}\dot{\phi}_i(t)\end{array}\right)
\end{equation}
where $m_i$ is the mass of the $i$th degree of freedom in the supersystem, $m_s$ is the mass of the system, and $\phi_i(t)$ is a function determined by the form of $\HSE$.  We can then write $\T_p$ in $2\times 2$ block form:
\begin{equation}
\T_p = \left(\begin{array}{cccc}\M_0 & \M_1 & \M_2 & \cdots \\ \\
		\begin{array}{cc}0&0\\0&0\end{array} &	
		\begin{array}{cc}1&0\\0&1\end{array} &	
		\begin{array}{cc}0&0\\0&0\end{array} &	 \cdots \\
		\vdots & \vdots & \vdots & \ddots
	\end{array}\right)\label{TpMatrixForm}
\end{equation}
In order to calculate $\T_p^{-1}$, we define $D_{ij}$ as the determinant of the $2\times 2$ sub-matrix $(\T_p)_{[1,2],[i,j]}$.  Then $\T_p^{-1}$ can be written simply as
\begin{equation}
\T_p^{-1} = \frac{1}{D_{12}}\left(\begin{array}{ccccc}\dot{\phi}_0 & -\frac{1}{m_s}\phi_0
	& D_{23} & D_{24} & \cdots \\ \\
	-m_s\ddot{\phi}_0 & \dot{\phi}_0 & -D_{13} & -D_{14} & \cdots \\
	0 & 0 & D_{12} & 0 & \cdots \\
	0 & 0 & 0 & D_{12} & \cdots \\ 
	\vdots & \vdots & \vdots & \vdots & \ddots\end{array}\right)
\end{equation}
By explicit computation, we verify that $\dot{\T}_p\T_p^{-1}$ takes the form of equation (\ref{MatrixForm}), with the coefficients given by:
\begin{widetext}
\begin{eqnarray}
\omeff^2 &=& \frac{\ddot{\phi_0}^2-\dddot{\phi_0}\dot{\phi_0}}{\dot{\phi_0}^2-\ddot{\phi_0}\phi_0}
	\label{Coeffs1}\\
\gameff&=&\frac{\dot{\phi_0}\ddot{\phi_0}-\dddot{\phi_0}\phi_0}{\dot{\phi_0}^2-\ddot{\phi_0}\phi_0}
	\label{Coeffs2}\\
F_{yi} &=& \sqrt{m_sm_i}\left(\dddot{\phi_i}-\gameff\ddot{\phi_i}+\omeff^2\dot{\phi_i}\right)
	\label{Coeffs3}\\
F_{qi} &=& \sqrt{\frac{m_s}{m_i}}\left(\ddot{\phi_i}-\gameff\dot{\phi_i}+\omeff^2\phi_i\right)
	\label{Coeffs4}\\
\hbar^2 f_1 &=&\sum_{i=1}^N{\sqrt{\frac{m_s}{m_i}}\Tr\left[\left(\begin{array}{cc}m_iF_{yi}\ddot{\phi}_i & m_iF_{qi}\ddot{\phi}_i\\
	F_{y_i}\dot{\phi}_i & F_{qi}\dot{\phi}_i\end{array}\right)\Ve\right]}\label{Coeffs5}\\
\hbar^2 f_2 &=&\sum_{i=1}^N{\sqrt{\frac{m_s}{m_i}}\Tr\left[\left(\begin{array}{cc}m_iF_{yi}\dot{\phi}_i & m_iF_{qi}\dot{\phi}_i\\
	F_{y_i}\phi_i & F_{qi}\phi_i\end{array}\right)\Ve\right]}\label{Coeffs6}
\end{eqnarray}
\end{widetext}
	
\subsection{Master Equation for the Inverted Harmonic Oscillator}

We begin with the supersystem Hamiltonian:
\begin{equation}
H = \frac{p^2}{2m_s} + \frac{m_s\Omega^2}{2}x^2 + \frac{q^2}{2m_e} - \frac{m_e\Lambda^2}{2}y^2
	+\alpha\sqrt{m_sm_e}xy
\end{equation}
The time translation matrix $\T$ is obtained by diagonalizing the equations of motion, which yields two normal modes.  This transformation is characterized by a new harmonic frequency $\omega$, a new inverse frequency $\lambda$, and a mixing angle \nolinebreak $\theta$:
\begin{eqnarray}
\omega^2 &=& \frac12\left(\Omega^2-\Lambda^2+
        \sqrt{\left(\Omega^2+\Lambda^2\right)^2+4\alpha^4}\right)\\
\lambda^2 &=& \frac12\left(\Lambda^2-\Omega^2+
        \sqrt{\left(\Omega^2+\Lambda^2\right)^2+4\alpha^4}\right)\\
\tan{\theta} &=& \frac1{2\alpha^2}\left(\Omega^2+\Lambda^2-
        \sqrt{\left(\Omega^2+\Lambda^2\right)^2+4\alpha^4}\right)
\end{eqnarray}
Using these quantities, the $\T_p$-matrix can be expressed in the form of equation (\ref{TpMatrixForm}), where the $\phi_i(t)$ are given by:
\begin{eqnarray}
\phi_0(t) &=& \cnt{2}\frac{\swt}{\omega}+\snt{2}\frac{\shlt}{\lambda}\label{Eqphi0}\\
\phi_1(t) &=& \frac{\sin{2\theta}}{2}\left(\frac{\swt}{\omega}-\frac{\shlt}{\lambda}\right)\label{Eqphi1}
\end{eqnarray}
Thus, for this case we immediately obtain the parameters of the master equation by direct substitution into equations (\ref{Coeffs1}-\ref{Coeffs6}).  All terms share a common denominator, which we denote by $D$:
\begin{widetext}
\begin{eqnarray}
D &=& \Denom \label{IHO_D}\\
\omeff^2 &=& \frac{\omega\lambda}{D}\left(\omega^2\cnt4-\lambda^2\snt4+
		\frac{\sin^2{2\theta}}{4}\left[\left(\omega^2-\lambda^2\right)\cwt\chlt-2\omega\lambda\swt\shlt\right]
		\right) \label{IHO_omeff} \\
\gameff &=& \frac{\left(\omega^2+\lambda^2\right)\sin^2{2\theta}}{4D}
		\left[\lambda\swt\chlt-\omega\cwt\shlt\right] \label{IHO_gameff}\\
F_y &=& \frac{-\sqrt{m_sm_e}\omega\lambda\left(\omega^2+\lambda^2\right)\sin{2\theta}}{2D}
		\left(\cnt2\chlt+\snt2\cwt\right) \label{IHO_Fy} \\
F_q &=& -\sqrt{\frac{m_s}{m_e}}\frac{\left(\omega^2+\lambda^2\right)\sin{2\theta}}{2D}
		\left(\omega\cnt2\shlt+\lambda\snt2\swt\right) \label{IHO_Fq}
\end{eqnarray}
The diffusion coefficients are naturally described in terms of the elements of the tensors in equations (\ref{Coeffs5}-\ref{Coeffs6}).  Factoring out their common prefactor $\beta=\frac{m_s}{4\hbar^2D}\sin^2{2\theta}\left(\omega^2+\lambda^2\right)$,
we obtain:
\begin{eqnarray}
f_1^{yy} &=& m_e\omega\lambda\beta
	      \left(\cnt2\chlt+\snt2\cwt\right)\left(\lambda\shlt+\omega\swt\right) \label{IHO_f1yy} \\
f_1^{yq} &=& \omega\lambda\beta\left(\cnt2\chlt+\snt2\cwt\right)\left(\chlt-\cwt\right) \label{IHO_f1yq} \\
f_1^{qy} &=& \beta\left(\omega\cnt2\shlt+\lambda\snt2\swt\right)\left(\lambda\shlt+\omega\swt\right)\label{IHO_f1qy}\\
f_1^{qq} &=& \frac{\beta}{m_e}\left(\omega\cnt2\shlt+\lambda\snt2\swt\right)\left(\chlt-\cwt\right)\label{IHO_f1qq}	\\
\nonumber\\
f_2^{yy} &=& m_e\omega\lambda\beta
	      \left(\cnt2\chlt+\snt2\cwt\right)\left(\chlt-\cwt\right)\label{IHO_f2yy} \\
f_2^{yq} &=& \beta\left(\cnt2\chlt+\snt2\cwt\right)\left(\omega\shlt-\lambda\swt\right)\label{IHO_f2yq} \\
f_2^{qy} &=& \beta\left(\omega\cnt2\shlt+\lambda\snt2\swt\right)\left(\chlt-\cwt\right)\label{IHO_f2qy}\\
f_2^{qq} &=& \frac{\beta}{m_e\omega\lambda}
	      \left(\omega\cnt2\shlt+\lambda\snt2\swt\right)\left(\omega\shlt-\lambda\swt\right)\label{IHO_f2qq}	
\end{eqnarray}
\end{widetext}

\section{Results and Analysis}

Having obtained a master equation describing the evolution of a system coupled to an IHE (inverted harmonic environment), we now proceed to examine the consequences of that evolution.  We have two tools for this analysis: on one hand, the master equation and its coefficients determine the instantaneous effects of the environment; on the other hand, we can explicitly evolve an initial state of the system to see the time evolution of state properties such as entropy and energy.  The master equation itself divides naturally into two parts corresponding to the two lines of equation (\ref{MasterEquation}); the terms in the first line produce renormalized unitary evolution (including external damping and forcing terms, which break unitarity), while the last two terms are diffusive and responsible for decoherence.  We thus divide our analysis into three sections, addressing in turn the quasi-unitary portion of the master equation, the diffusive terms in the master equation, and the behavior of evolved observables.  Our primary results are in equations (\ref{ApproxS_of_t}-\ref{Thumbnail}), where we demonstrate the linear growth of entropy in the system at the rate set by the Lyapunov exponent and obtain an approximate decoherence timescale that turns out to be \emph{logarithmically} dependent on the coupling.  This implies that isolation from chaotic environments is in a sense exponentially difficult.  In particular, it is even harder to isolate a system from a chaotic environment than the many harmonic oscillators of the QBM environment, where the decoherence time is approximately quadratic in the coupling strength \cite{UnruhZurek89PRD,HuPazZhang92PRD}.  Thus, the reader who wishes to skip straight to the discussion of the implications of quantum chaos for decoherence may skim most of sections \ref{UnitaryDiscussion} and \ref{DiffusionDiscussion}, which analyze the master equation in detail.

\subsection{Unitary evolution}
\label{UnitaryDiscussion}

The first five terms in the right-hand side of equation (\ref{MasterEquation}),
\begin{widetext}
\begin{equation}
        \frac{1}{i\hbar}\left(\frac{m\omeff^2}{2}\commutator{\hat{x}^2}{\hat{\rho}}
        +\frac{1}{2m}\commutator{\hat{p}^2}{\hat{\rho}}
        +\frac{\gameff}{2}\commutator{\hat{x}}{\anticommutator{\hat{p}}{\hat{\rho}}}
        -F(t)\commutator{\hat{x}}{\hat{\rho}}\right)
	\label{UnitaryTerms}
\end{equation}
\end{widetext}
are exactly those for the evolution of an isolated harmonic oscillator subject to an external force $F(t)$ and a damping force $\gameff(t)$.  For convenience, although the term $\gameff\commutator{\hat{x}}{\anticommutator{\hat{p}}{\hat{\rho}}}$ breaks unitarity, we refer to these terms in the master equation as the ``unitary evolution'' terms.

\begin{figure*}
\includegraphics[width=6.4in]{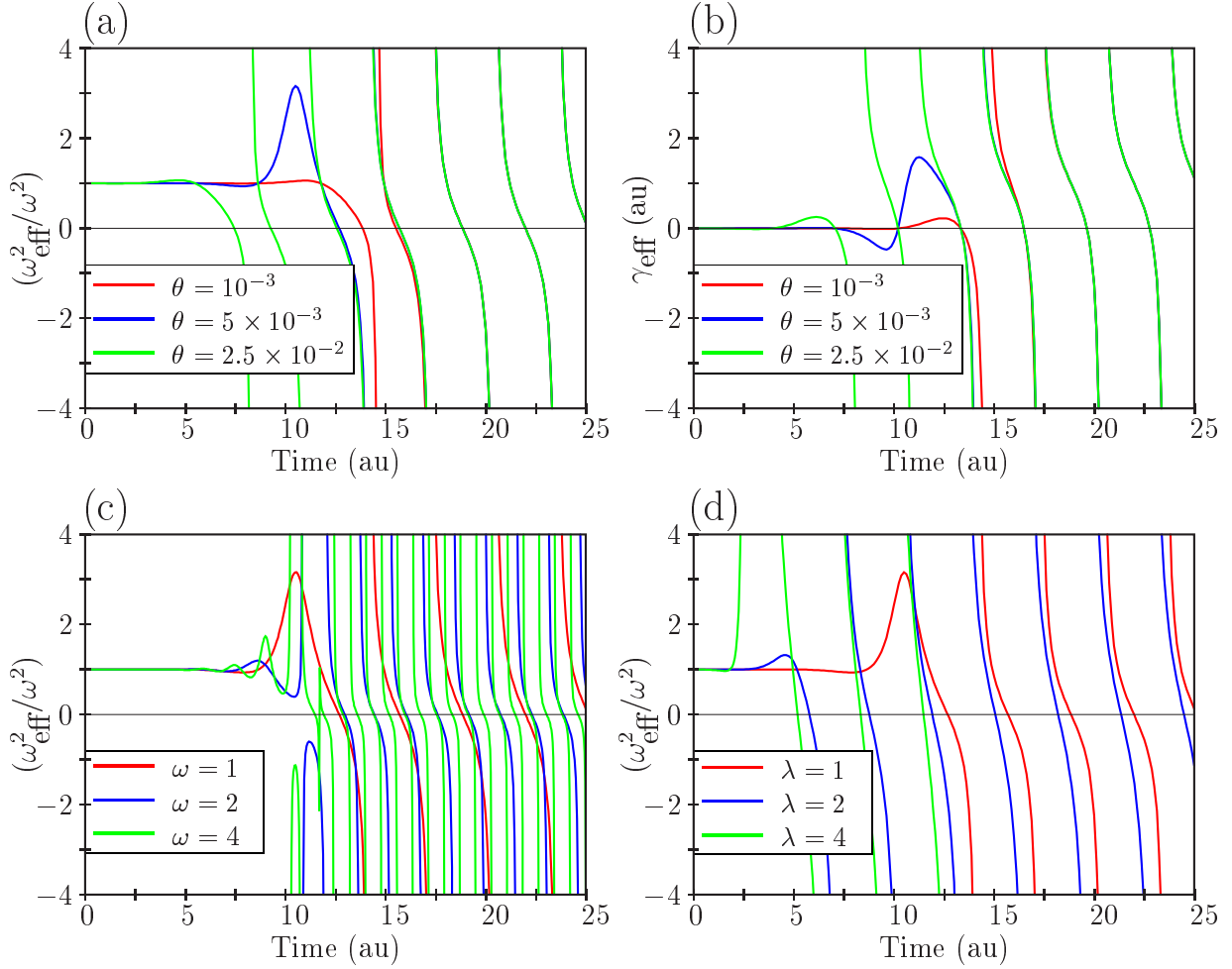}
\caption{Dependence of the master equation coefficients $\omeff^2$ and $\gameff$ on coupling angle $\theta$, system frequency $\omega$, and the effective Lyapunov exponent $\lambda$ of the environment.  The base configuration is $\omega=\lambda=m_s=m_e=1$, $\theta=10^{-3}$: plot (a) shows the dependence of $\omeff^2(t)$ on $\theta$, plot (b) shows the dependence of $\gameff(t)$ on $\theta$, plot (c) shows the dependence of $\omeff^2(t)$ on $\omega$, and plot (d) shows the dependence of $\omeff^2(t)$ on $\lambda$.  The units of time in all plots are identical but arbitrary.}
\label{figU}
\end{figure*}

Because the coefficients of all the terms except $\commutator{\hat{p}^2}{\hat{\rho}}$ are time-dependent, however, the evolution induced by these terms is not necessarily intuitive.  The time-dependence of $\omeff^2$ and $\gameff$ (for various values of the parameters) is shown in Figure \ref{figU}.   We see immediately that although the coefficients are \emph{initially} very close to their bare values ($\omeff^2 \sim \omega^2$, $\gameff \sim 0$), they begin after a certain time to vary dramatically, and finally appear to converge to a periodically diverging function of time.  Thus, coefficients that we expect to be positive and well-defined (and which, in the QBM model, settle down after a while to stable values) not only take on negative values, but appear at certain times to become infinite.

The root of this behavior lies in the common denominator of all the coefficients (equation (\ref{IHO_D})).  This denominator, the determinant of the upper-left $2\times 2$ sub-matrix of $\T_p$, takes on both positive and negative values -- and thus passes through zero.  Since the various numerators in equations (\ref{IHO_omeff}-\ref{IHO_f2qq}) do not change sign in sync with the denominator, all the coefficients not only vary from positive to negative, but diverge whenever $D=0$.  While this phenomenon is unexpected and counter-intuitive, we note that it does \emph{not} indicate unphysical behavior.  Detailed analysis (omitted here) shows that near a divergence in the coefficients, the system state $\rho$ is forced to assume a form such that the effects of the superoperators in the master equation cancel each other out; thus, when the coefficients diverge, their effects sum to a perfectly finite value, and the evolution of the state is completely physical.

We conclude from this that the divergences are not symptomatic of any physical phenomenon, but are rather a consequence of the system-environment paradigm that we have imposed on the supersystem.  As the system and environment interact with each other, information about the initial $x_0$ and $p_0$ of the system is transferred to the environment (and vice-versa).  At certain times, \emph{all} information about a particular linear combination of $x_0$ and $p_0$ has been transferred to the environment, in return for which all information about some linear combination of $y_0$ and $q_0$ resides in the system.  This is true only for an instant, but at that instant $\hat{\rho}(t)$ does not uniquely determine $\hat{\rho}(0)$, no differential equation for $\hat{\rho}$ can exist, and the master equation necessarily breaks down.  The mathematical reflection of this is that the upper left $2\times 2$ sub-matrix of $\T_p$ (which specifies the relationship between $(x,p)$ and $(x_0,p_0)$) is instantaneously non-invertible, because $D=0$.

For $t\sim 0$, $D=1$.  Assuming that the coupling is weak ($\theta \ll 1$), we can expand $D$ for $\lambda^{-1}\ll t \ll -2\lambda^{-1}\log\theta$ as
\begin{equation}
D \simeq 1 + \theta^2 e^{\lambda t} \frac{\left(\omega^2-\lambda^2\right)\swt+2\omega\lambda\cwt}{\omega\lambda}.
\label{ApproxD}
\end{equation}
Since the fraction at the end of equation (\ref{ApproxD}) is $O(1)$, we conclude that $D\simeq 1$ until shortly before the first divergence occurs; further, the timescale of that divergence is
\begin{equation}
t_c \simeq -2\frac{\log\theta}{\lambda} + \log{\left(\frac{\omega\lambda}{\omega^2+\lambda^2}\right)}.
\label{CriticalTime}
\end{equation}
While the first divergence may occur later than $t_c$, it cannot occur earlier.  Thus, we have a natural time scale in the system that divides time into two regions: an initial period, when $t<t_c$; and long times, when $t>t_c$.  In the initial period, the unitary coefficients of the master equation are well-approximated by their bare values, whereas in the long-time regime the coefficients' values periodically diverge and are only indirectly related to their bare values.  When $t \simeq t_c$, the coefficients' values are difficult to characterize.  Finally, we note that the critical timescale is reflected in the diffusive coefficients as well (see Figure \ref{figD}), but \emph{not} necessarily in physical quantities obtained from $\hat{\rho}(t)$ itself, as we expect from the argument that the divergences are not reflected in physical quantities.  

Nonetheless, $t_c$ is a useful time scale to keep in mind because in the initial period we can be assured that the renormalized unitary evolution is very similar to the bare unitary evolution; thus, the effects of interaction with the environment during this regime will be all but unnoticeable on classical scales.  Beyond $t_c$, the terms in the master equation that govern the evolution of large-scale structures in phase space \emph{may} depart dramatically from their uncoupled values; we cannot be certain.  If we take as the criterion for our model's relevance to a real chaotic environment that its effects be small on classical scales, then relevance is guaranteed for the initial period, but not for long times.  Thus, although we will examine the long time behavior of the model at times, we regard the initial ($t<t_c$) period as a candidate model of decoherence due to a chaotic environment.

Another perspective on the same conclusion comes from the fact that any physical chaotic environment will have a bounded phase space.  The operator $\envop$ that couples to a system operator $\sysop$ will thus have a bounded spectrum, and the expectation value of the coupling $\Hint = \sysop\envop$ will also be bounded by some value $\expect{\Hint}_{\text{max}}$.  As long as our model respects this constraint, we can consider it a plausible model for that environment; however, because the $\hat{y}$ operator is unbounded and $\expect{\hat{y}^2}$ increases exponentially in time, our model will eventually expand into a larger phase space, and the coupling will dominate the overall Hamiltonian.  The critical time $t_c$ indicates roughly when $\Hint$ begins to dominate the overall Hamiltonian.

For completeness, we consider briefly the effects of an environment which is \emph{not} unstable -- a simple harmonic oscillator or free particle.  The simple harmonic environment does not have divergences in the master equation coefficients.  Instead, $\omeff^2$ and $\gameff$ oscillate stably around $\omega^2$ and $0$ (respectively); $\omeff^2$ is plotted in Figure \ref{figSHO}.  The free particle is somewhat more interesting; a wave packet spreads slowly ($\Delta x^2 \propto t$) in the absence of a potential, so in this sense there is a weak irreversibility.  We'll discuss this in more detail when we consider entropy, but as Figure \ref{figSHO} shows, the divergences characteristic of the IHE \emph{do} plague the free-particle environment.  However, $t_c$ is no longer logarithmic in the coupling, but rather obeys a power-law; thus the free-particle environment can be more effectively isolated from the system (like the SHO environment) by making the coupling strength very small.
 
\subsection{Diffusive terms}
\label{DiffusionDiscussion}

The last two terms in equation (\ref{MasterEquation}),
\begin{equation}
- f_1(t)\commutator{\hat{x}}{\commutator{\hat{x}}{\hat{\rho}}}
+ f_2(t)\commutator{\hat{x}}{\commutator{\hat{p}}{\hat{\rho}}},
\end{equation}
are non-unitary and diffusive, and produce decoherence.  If $\hat{\rho}$ is transformed to a Wigner function $W(x,p)$, then equation (\ref{MasterEquation}) becomes a Fokker-Planck type equation.  The $f_1$ or ``normal diffusion'' term is seen to produce diffusion in momentum space according to $\dot{W} \propto -f_1\pdiff{^2}{^2p}W$, while the $f_2$ or ``anomalous diffusion'' term tends to skew the state according to $\dot{W} \propto -f_2\pdiff{}{x}\pdiff{}{p}W$ (see \cite{UnruhZurek89PRD}, where ``anomalous'' diffusion was first identified).  This is also apparent in equations (\ref{MEdxdt}-\ref{MEdxpdt}), where $f_1$ appears only in $\pdiff{\expect{\hat{p}^2}}{t}$ while $f_2$ appears only in $\pdiff{\expect{\anticommutator{\hat{x}}{\hat{p}}}}{t}$.

\begin{figure*}
\includegraphics[width=6.4in]{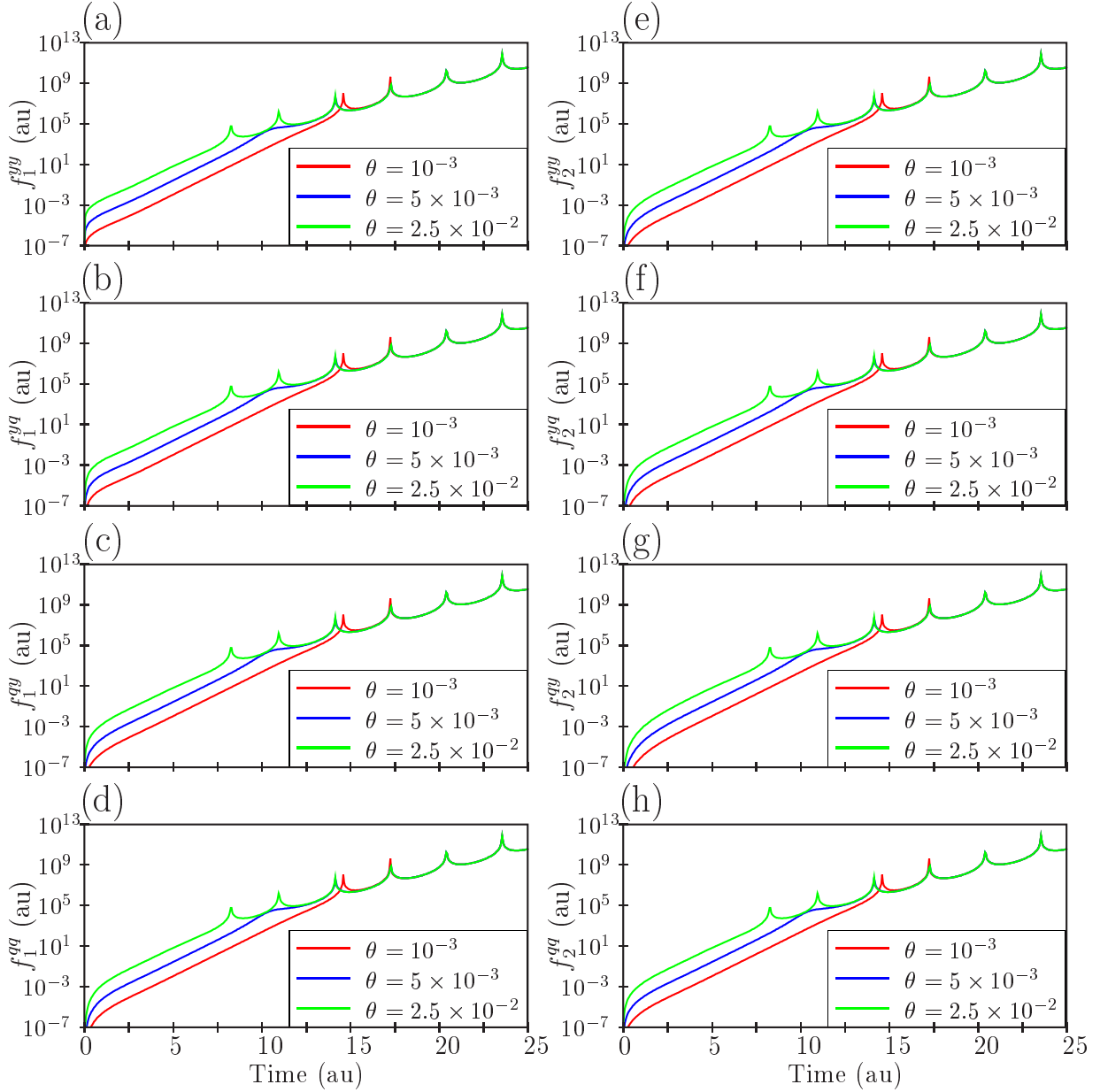}
\caption{The absolute values of the diffusive master equation coefficients $f_1$ and $f_2$ as a function of time, plotted on a logarithmic scale in arbitrary units for three different values of the coupling angle $\theta$.  Since $f_1$ and $f_2$ are dependent on the state of the environment, the four tensor components of each $f_n$ are plotted here (note, however, that all eight components are nearly identical in this case).  Plot series (a-d) shows the tensor components of $f_1$, while plot series (e-h) shows the tensor components of $f_2$.  The base configuration is $\omega=\lambda=m_s=m_e=1$.  We emphasize that the coefficients actually change sign regularly (at each cusp, in fact); in order to use a logarithmic scale, the absolute values must be plotted in place of the signed values.}
\label{figD}
\end{figure*}

\begin{figure*}
\includegraphics[width=6.4in]{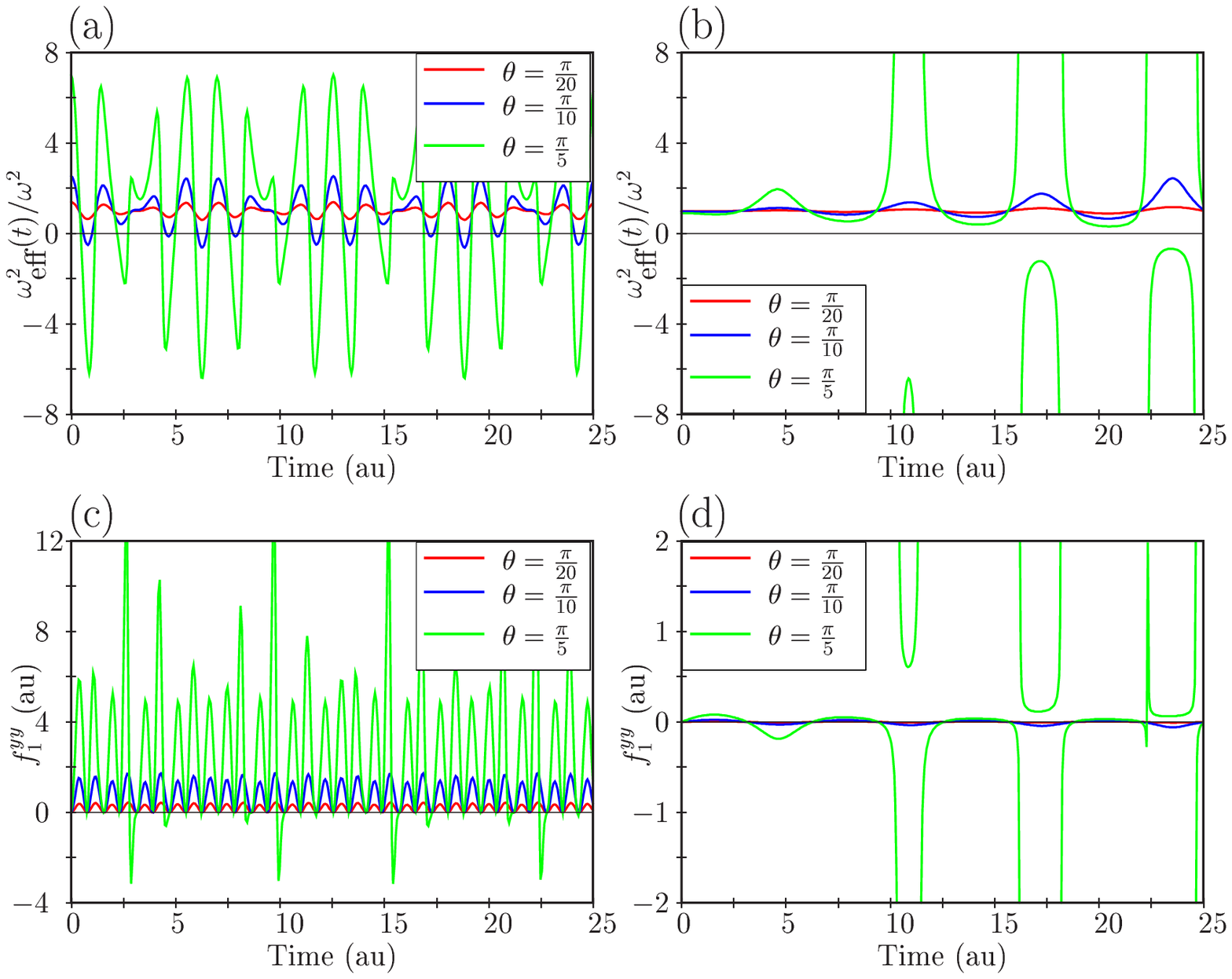}
\caption{Selected master equation coefficients indicative of the behavior of harmonic-oscillator (plots (a) and (c)) and free-particle (plots (b) and (d)) environments are plotted.  We plot $\omeff^2/\omega^2$ and $f_1^{yy}$ for imaginary $\lambda$ (that is, the environment is a single SHO with frequency $-i\lambda$) and for $\lambda=0$.  The base configuration is $\omega=\frac12$, $-i\lambda=4$, $\theta=\frac{\pi}{10}$, and $m_s=m_e=1$.  Plot (a) shows $\omeff^2/\omega^2$ for $\lambda=4i$; plot (b) shows $\omeff^2/\omega^2$ for $\lambda=0$; plot (c) shows $f_1^{yy}$ for $\lambda=4i$; and plot (d) shows $f_1^{yy}$ for $\lambda=0$.  We emphasize that the coupling ($\theta$) is much greater than in the cases examined previously, yet there are no divergences in $\omeff^2$ or $\gamma$ for the SHO environment, and the diffusion coefficients (represented by $f_1^{yy}$) remain relatively small and bounded.  Divergences do appear in the coefficients for the free-particle environment, but their onset time is polynomial in $\theta^{-1}$, and they do not appear in these plots except for the largest value of $\theta$ ($\pi/5$).}
\label{figSHO}
\end{figure*}

In QBM, the coefficients $f_1$ and $f_2$ of these terms equilibrate to constant values or monotonically decreasing functions after an initial period; in our IHE model these coefficients, like those of the unitary terms, vary widely over time.  Unlike the other coefficients, $f_1$ and $f_2$ are dependent on the state of the environment; however, they can each be written as the contraction of a $2\times 2$ coefficient tensor with the variance tensor of the environment at $t=0$ (see equations (\ref{IHO_f1yy}-\ref{IHO_f2qq})):
\begin{equation}
f_n(t) = \Tr\left[\left(\begin{array}{cc}f_n^{yy}&\frac12 f_n^{yq}\\ \frac12 f_n^{qy} & f_n^{qq}\end{array}\right)
	\left(\begin{array}{cc}\Delta y^2 & \Delta yq \\ \Delta yq & \Delta q^2 \end{array}\right)\right].
\end{equation}
The absolute values of the four subcoefficients for each of $f_1$ and $f_2$ are plotted in Figure \ref{figD}, for the same base parameters as Figure \ref{figU} and for three different coupling strengths.  Each subcoefficient has a prefactor involving the masses of the system and environment, which is ignored in these plots; thus, if the system and environment masses are substantially different, one subcoefficient may be promoted over another.  The most salient feature of these plots is that they appear identical -- only at times shorter than $\lambda^{-1}$ is any difference visible at all between the various coefficients.  Examination of equations (\ref{IHO_f1yy}-\ref{IHO_f2qq}) confirms that for $t\gg\lambda^{-1}$, all the subcoefficients are proportional to $\theta^2 D^{-1}e^{2\lambda t}$.  This also explains the sharp distinction between short- and long-time behavior in Figure \ref{figD}.  In the short-time regime, $D\simeq 1$, and each subcoefficient is well-approximated by $f\propto e^{2\lambda t}$; in the long-time regime, however, $D\simeq \theta^2 e^{\lambda t} \cos(\omega t + \phi)$, and $f\propto e^{\lambda t}\sec(\omega t + \phi)$.  Thus, on the log-plot of Figure \ref{figD}, we see two distinct lines with slopes $2\lambda$ and $\lambda$, the second of which is punctuated by the periodic divergences seen in all the master equation coefficients.  In order to make this explicit, we examine the short-time ($\lambda^{-1}<t<t_c$) behavior of $f_1$, which is responsible for entropy production (see section \ref{results_SandE}), when the mass ratio $m_{\text{env}}/m_{\text{sys}} \ll 1$.  In this limit, $f_1$ is dominated by $f_1^{qq}$; using the approximations $\shlt\gg 1$, $\theta\ll 1$ and $D\simeq 1$, we obtain:
\begin{equation}
\hbar^2 f_1 \simeq \frac{m_s\theta^2(\omega^2+\lambda^2)}{m_e}e^{2\lambda t}
\label{Approxf1}
\end{equation}
The key point in equation (\ref{Approxf1}) is that the coefficient of the term that produces diffusion in momentum (the primary factor in decoherence) increases exponentially with time and $\lambda$, but is only quadratic in the coupling strength.

The change in the exponent of the $f$ coefficients at $t\sim t_c$ is physically relevant as well as mathematically sensible.  In the short-time regime, the oscillatory dynamics of the system and the hyperbolic stretching of the environment proceed largely independently of one another; just as the environment induces only minor perturbations in the system, the system does not disturb the environment greatly.  Thus, the stretching of the environment along its unstable manifold is reflected in the system as diffusion in one of the two phase-space dimensions.  After $t_c$, however, the interaction Hamiltonian begins to dominate the dynamics of the system, the overall dynamics is strongly coupled, and the unstable manifold of the environment rotates.  Diffusion in the system is averaged over stable and unstable directions, and so the diffusion coefficients increase only as $e^{\lambda t}$.  Nonetheless, because diffusion now occurs along all directions in phase space, the entropy of $\hat{\rho}$ continues to grow at the same rate (as will be shown in section \ref{results_SandE}).

\subsection{Entropy of the Reduced Density Matrix}
\label{results_SandE}
Having analyzed in detail the IHE master equation, we turn finally to the actual behavior of $\hat{\rho}(t)$.  Because decoherence manifests as entropy production in $\hat{\rho}$, we first calculate the Von Neumann entropy $\entropy(t)=\Tr{(\hat{\rho}\ln\hat{\rho})}$ of the reduced density matrix, and examine the dependence of $\entropy(t)$ on various parameters.  However, entropy production can reflect not only the destruction of quantum coherences but also the destruction of large-scale ``classical'' structures in phase space.  In order to verify the ability of the IHE to produce decoherence \emph{without} destroying classical structures, we take the expectation value of energy $\energy=\expect{\HS}$ as a convenient classical quantity, and show that for appropriate initial conditions a dramatic rise in $\entropy$ occurs while $\energy$ remains relatively undisturbed.  It is worth noting that because all the states we consider have $\expect{\hat{x}}=\expect{\hat{p}}=0$, this is probably an overly strong condition, since large (on classical scales) amounts of energy can be added to the system by a simple Galilean transformation at $t=0$, which changes nothing of the analysis except for adding a constant offset to $\energy(t)$.

\subsubsection{Entropy}

The canonical state to examine is a ``Schr\"odinger Cat'' state, typically a superposition of widely separated coherent states.  We have examined the effects of the IHE on such a state, but because the state is not itself Gaussian the analysis is quite messy in our ansatz, and so we will discuss instead the behavior of a squeezed Gaussian state that is highly extended in $x$.  With respect to a diffusive decoherence process, the most significant features of a ``cat'' state are the interference fringes in $W(x,p)$ that lie between the two Gaussian bumps.  They are extended in $x$ but equivalently narrow in $p$.  A Gaussian state which is highly squeezed in momentum has similar small-scale structure, reflected in the strong off-diagonal correlations of $\hat{\rho}$.  For convenience, we consider such states.

In general, $\Tr{(\hat{\rho}\log\hat{\rho})}$ is difficult to compute.  For Gaussian states, however, $\entropy$ can be easily computed in terms of the state's scaled area in phase space.  The Heisenberg uncertainty relation $\Delta x \Delta p \ge \frac{\hbar}{2}$ provides a fundamental unit of phase space area.  We thus define $a_0 = \frac{\hbar}{2}$, and define the phase space area occupied by a Gaussian state as
\begin{equation}
a = \sqrt{\Delta x^2\Delta p^2 - \left(\Delta xp\right)^2},
\end{equation}
and the scaled area as $A=a/a_0$.  See \ref{CumulantNote} after equation \ref{EnvVarianceTensor} for an explanation of the $\left(\Delta xp\right)$ notation.  Every Gaussian pure state has an area $a=a_0$; mixed states have $a>a_0$.  It can be shown \cite{ZurekHabibPaz93PRL} that the entropy is given exactly by
\begin{equation}
\entropy = \frac12\left[(A+1)\ln(A+1)-(A-1)\ln(A-1)\right]-\ln(2).
\label{ExactEntropy}
\end{equation}
A convenient approximation, which is exact for $\entropy=0$ and is always accurate to within $1-\ln2 \simeq 0.31$, is
\begin{equation}
\varentropy = \ln(A) = \frac12\ln(A^2),
\label{ApproxEntropy}
\end{equation}
where the second equality is useful because $A^2$ is easily calculable as the determinant of the state's variance tensor.  This quantity is related to (but not identical to!) the linear entropy $\varsigma=1-\Tr\rho^2$ because $\Tr\rho^2 = A^{-1}$.  Thus, the linear entropy, $\varsigma=1-A^{-1}$, is a first order (in $A^{-1}$) approximation to $\varentropy = \ln(A)$.

\subsubsection{Analysis of the $\entropy(t)$ Plots}

\begin{figure*}
\includegraphics[width=6.4in]{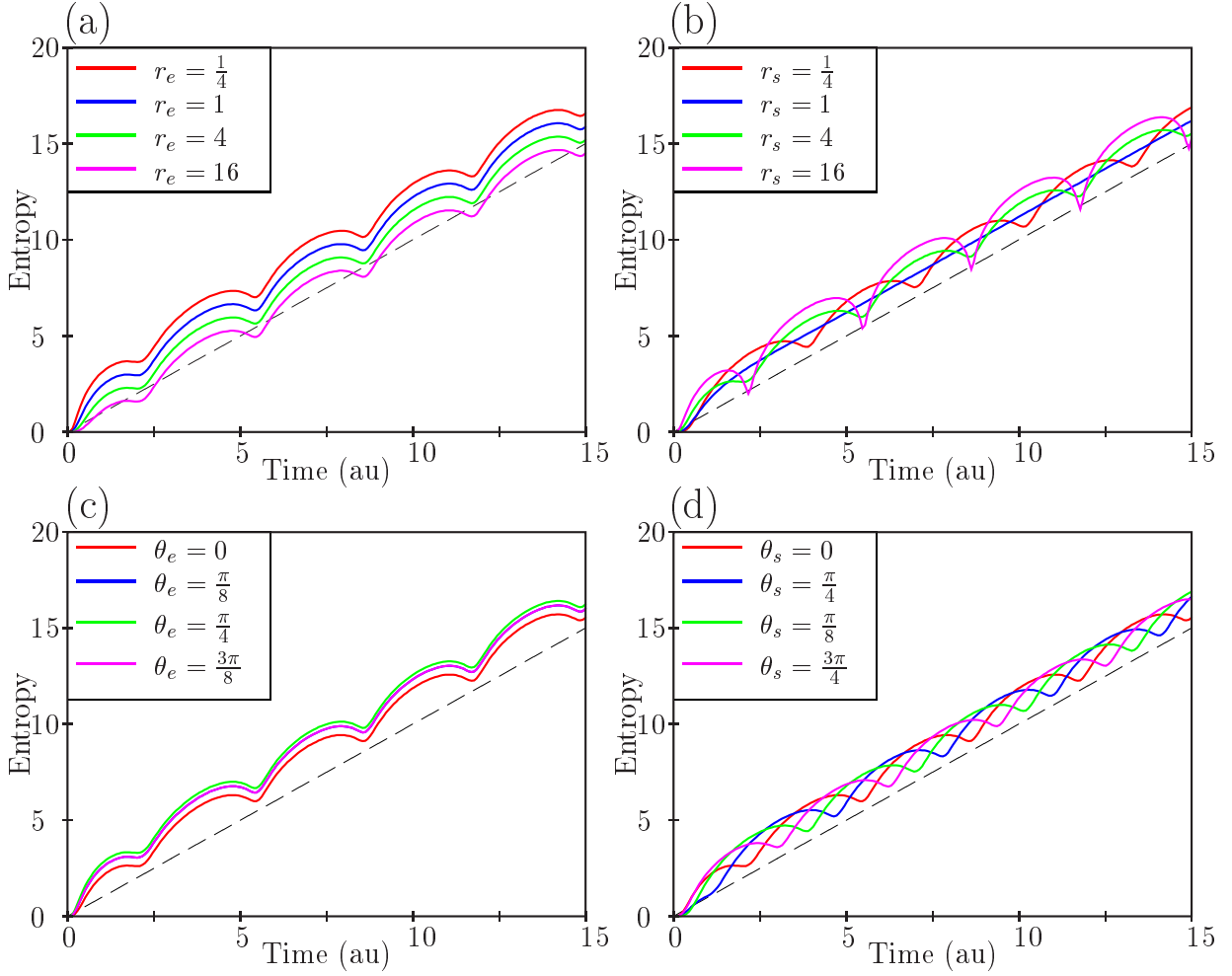}
\caption{Dependence of system entropy on the initial states of the system and the environment.  The base parameters are $\omega=\lambda=m_s=m_e=1$, $\theta=\frac{\pi}{64}$, $r_s=4$, $r_e=2$, $\theta_s=0$, and $\theta_e=0$.  Plot (a) shows the dependence of $\entropy(t)$ on $r_e$, the squeezing parameter of the environment; plot (b) shows the dependence of $\entropy(t)$ on $r_s$; plot (c) shows the dependence of $\entropy(t)$ on $\theta_e$, the squeezing angle of the environment; and plot (d) shows the dependence of $\entropy(t)$ on $\theta_s$.  All entropies are Von Neumann entropies: $\entropy=\mbox{Tr}\rho\ln\rho$, and thus are dimensionless.}
\label{figS1}
\end{figure*}

\begin{figure*}
\includegraphics[width=6.4in]{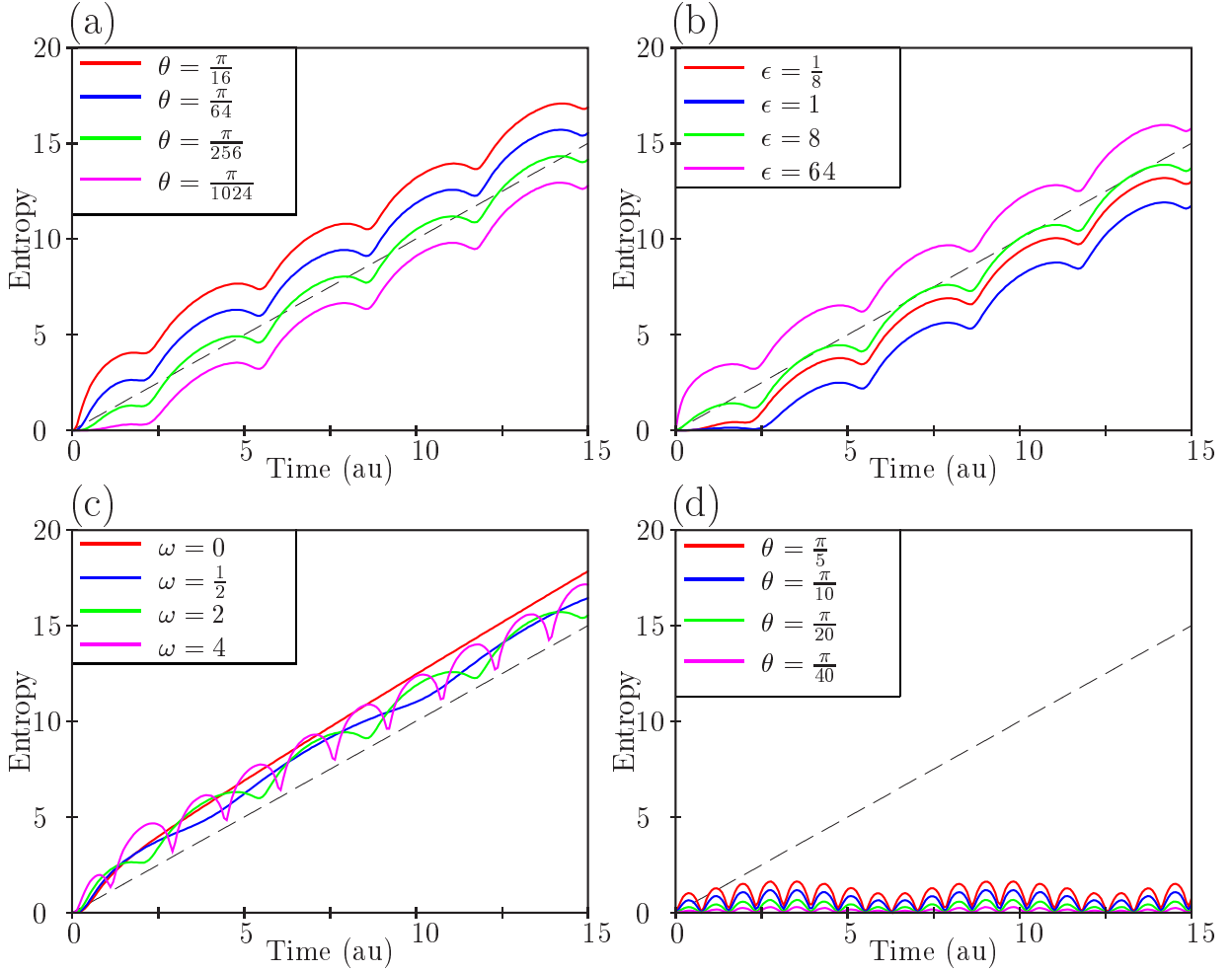}
\caption{Dependence of system entropy on the parameters of the master equation.  The base parameters for plots (a-c) are $\omega=\lambda=m_s=m_e=1$, $\theta=\frac{\pi}{64}$, $r_s=4$, $r_e=2$, $\theta_s=0$, and $\theta_e=0$; for plot (d) the parameters are $\lambda=4i$, $\omega=m_s=m_e=1$, $\theta=\frac{\pi}{64}$, $r_s=4$, $r_e=2$, $\theta_s=0$, and $\theta_e=0$.  Plot (a) shows the dependence of $\entropy(t)$ on $\theta$; plot (b) shows the dependence of $\entropy(t)$ on $m_s/m_e$; plot (c) shows the dependence of $\entropy(t)$ on $\omega$; and plot (d) shows the dependence of $\entropy(t)$ for a \emph{stable} environment on $\theta$.  We emphasize that in plot (d), entropy oscillates but does not increase over time (compare with figure \ref{figS3}).  The couplings are much larger for plot (d), in order to make $\entropy$ perceptible on the same scale as is used for the unstable environment.}
\label{figS2}
\end{figure*}

\begin{figure*}
\includegraphics[width=6.4in]{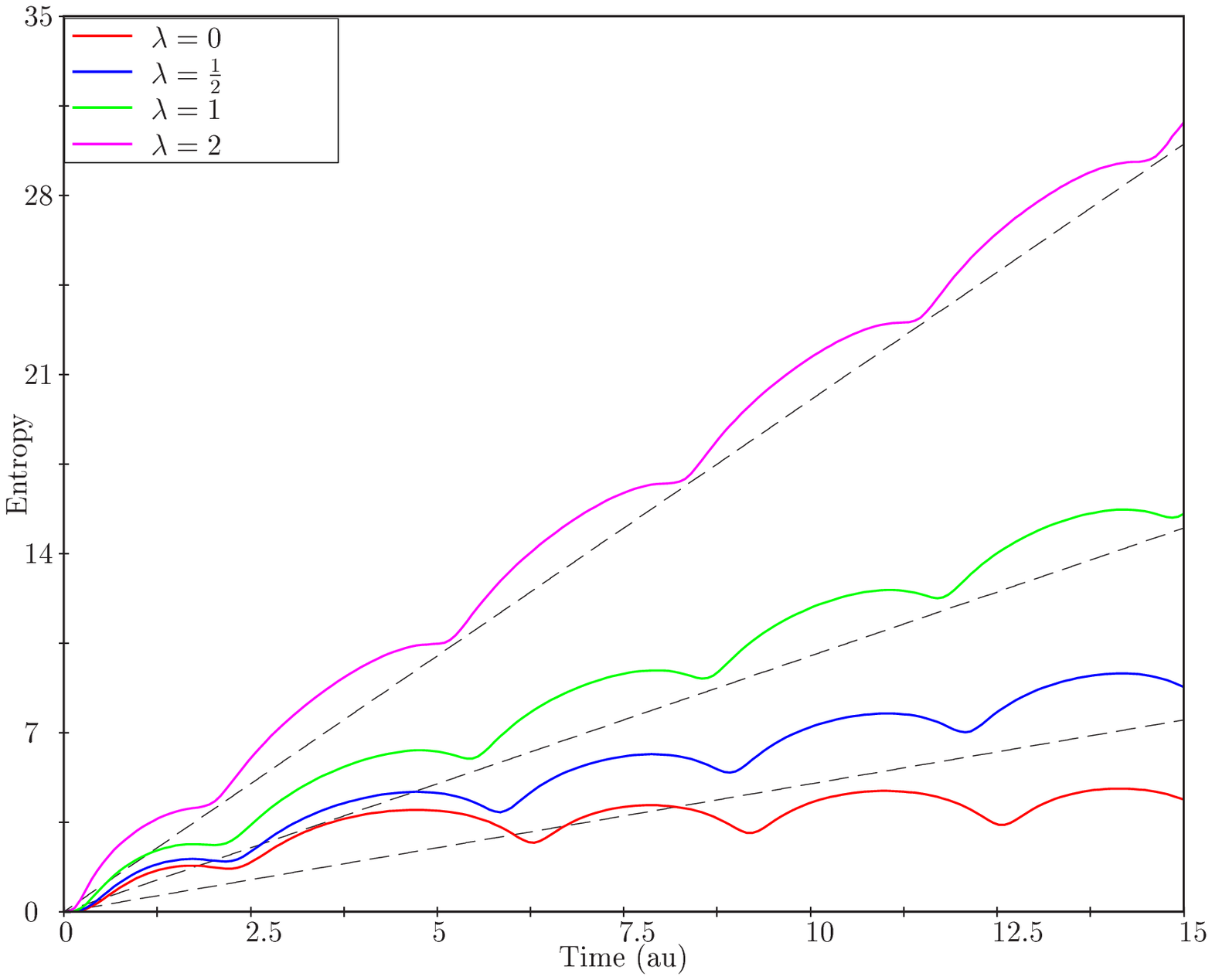}
\caption{Dependence of system entropy on the effective Lyapunov exponent.  The base parameters are $\omega=m_s=m_e=1$, $\theta=\frac{\pi}{64}$, $r_s=4$, $r_e=2$, $\theta_s=0$, and $\theta_e=0$.  The curve for $\lambda=0$ shows logarithmic increase in entropy.}
\label{figS3}
\end{figure*}

In Figures \ref{figS1}-\ref{figS3}, $\entropy(t)$ is plotted for a wide range of initial conditions and parameters.  The initial conditions consist of the squeezing parameter $r=\frac{\Delta x}{\Delta p}$ and squeezing angle $\theta$ for both the system and the environment (large $r$ and $\theta=0$ indicates a state extended in position and squeezed in momentum, while $\theta=\frac{\pi}{2}$ implies the reverse).  The parameters of the master equation include the bare $\omega$ and $\lambda$, the coupling $\theta$, and mass-ratio $\epsilon=m_e/m_s$.  The base parameters used in Figures \ref{figS1}-\ref{figS3} are: $\omega=1$, $\lambda=1$, $\theta=\frac{\pi}{64}$, $m_s=1$, $m_e=1$, $r_s=4$, $r_e=2$, $\theta_s=0$, and $\theta_e=0$.  Each of the plots varies one of these parameters, except for the fourth plot in Figure \ref{figS2}, in which the entropy for a \emph{non-inverted} harmonic environment is plotted for various values of the coupling angle.

All but one of the plots demonstrate the same basic behavior: entropy increases linearly as $\entropy = \lambda t + \entropy_0$, with periodic modulations.  The last plot in Figure \ref{figS2} shows $S(t)$ for a stable environment; the entropy oscillates in time but does not increase irreversibly.  Figure \ref{figS3} demonstrates the dependence on $\lambda$, with the intriguing caveat that for $\lambda=0$ we obtain not constant entropy, but logarithmically increasing entropy.  This is explained by the fact that for $\lambda=0$ the environment is a free particle, whose wave packet spreads as $\Delta y^2 \propto t$; this mild irreversibility produces entropy growth that is only asymptotically zero.  For other values of $\lambda$, $\Delta y^2 \propto e^{2\lambda t}$, and entropy grows linearly.

Varying the squeezing parameter or squeezing angle of the environment (Figure \ref{figS1}) changes only the initial jump in entropy, $\entropy_0$.  $\entropy_0$ is minimal for an unsqueezed environment ($r_e=1$), and increases approximately as $\log(r_e+r_e^{-1})$.  The dependence on $\theta_e$ is minimal for $\epsilon=1$; for larger or smaller mass ratios, the dependence on the squeezing angle becomes more noticeable.  It should be noted that the phrase ``initial jump in entropy'' refers precisely to the intercept of $\entropy(t) \simeq \lambda t + \entropy_0$ in the long time limit; $\entropy_0$ can be negative, which means only that the linear growth in entropy is postponed for a time $-\entropy_0/\lambda$.

Also in Figure \ref{figS1}, it is apparent that the effect of varying $r_s$ and $\theta_s$ is merely to modify the periodic modulation of $\entropy(t)$.  Since the system state rotates in phase space (according to the dynamics) over time anyway, changing $\theta_s$ merely changes the phase of the oscillation in $\entropy(t)$, while $r_s$ determines the shape and amplitude of the oscillation: it becomes more dramatic as $r_s$ increases or decreases from $1$ (which eliminates the modulation).

Figures \ref{figS2}-\ref{figS3} show the effects of varying $\omega$, $\lambda$, $\epsilon$, and $\theta$.  Both $\theta$ and $\epsilon$ affect $\entropy_0$ without changing the character of $\entropy(t)$ in any other way.  This is particularly interesting for the coupling $\theta$; regardless of how small the coupling is, the entropy still grows as $e^{\lambda t}$.  In particular, the plot for $\theta = \frac{\pi}{1024}$ corresponds to a crossover time of $t_c > 15$, yet entropy begins to grow linearly around $t \sim 2.5$.  Changing $\lambda$, as mentioned earlier, changes the rate of entropy production.  Finally, the result of varying $\omega$ is a change in the periodic modulation; the frequency of this modulation is of course $\omega$.  Of special interest is the curve for $\omega=0$; here, although the periodic modulation has vanished, $\entropy(t)$ rises not as ${\lambda t}$ but as ${\lambda t} + \log(\lambda t)$.  The explanation is the same as it was for the $\lambda=0$ case; when $\omega=0$ the system is a free particle, and the linear spread of $\Delta x^2$ contributes a logarithmic term to the entropy.

\subsubsection{Conclusions regarding Entropy and Energy}

The linear growth of entropy seen in Figures \ref{figS1}-\ref{figS3} indicates that the IHE model can continue to decohere a system to which it is coupled after other models, such as QBM, have ceased to produce substantial entropy.  However, it does not demonstrate that the IHE can produce decoherence \emph{faster} than QBM.  To examine this issue, we first derive a formula for the rate of entropy growth.

The exact entropy (equation (\ref{ExactEntropy})) is difficult to work with; therefore we approximate $\entropy(t)$ with $\varentropy(t) = \frac12\ln{A^2}$ from equation (\ref{ApproxEntropy}).  Using equations (\ref{MEdxdt}-\ref{MEdxpdt})), we obtain
\begin{equation}
\pdiff{}{t}(A^2) = -\gameff(t)A^2 + 2\left[f_1(t)\Delta x^2-f_2(t)(\Delta xp)\right].
\end{equation}
Not surprisingly, a positive dissipation coefficient $\gameff$ causes the state to shrink in phase space; however, as we saw in the analysis of the unitary terms, $\gameff$ can be both positive and negative, and thus expands the state as often as it shrinks it.  We conclude that the $\gameff$ term contributes only to the periodic modulation of $\entropy(t)$.  The effect of the $f_2$ term is highly dependent on $\Delta xp$, which can be positive, negative, or zero.  However, because $\Delta xp$ oscillates around $0$ as the system evolves, this term will contribute primarily to the periodic modulation as well.  This leaves:
\begin{equation}
\pdiff{}{t}(A^2) \simeq 2f_1(t)\Delta x^2
\label{ApproxA2dot}
\end{equation}
Since $\Delta x^2$ is strictly positive and $f_1(t)$ in the short-time regime $t<t_c$ is positive (see equation (\ref{Approxf1})), this predicts monotonic growth in $A^2$, and thus $\entropy(t)$.  In addition, we can use the previous analysis of $f_1(t)$ to approximate
\begin{equation}
f_1(t) \simeq \kappa^2\theta^2 e^{2\lambda t}
\end{equation}
(where $\kappa$ depends on $\omega$, $\lambda$, $\epsilon$, and the initial state of the environment, but not on time) and integrate equation (\ref{ApproxA2dot}), obtaining
\begin{equation}
\varentropy(t)) = \log{(\kappa\Delta x^2)}+\log{\theta}+\lambda t.
\label{ApproxS_of_t}
\end{equation}
This equation summarizes our most significant results.  The amount of entropy produced is linear in $t$ and in $\lambda$, but only logarithmically dependent on the properties of the initial state \emph{or} the coupling strength.  We can invert equation (\ref{ApproxS_of_t}) to obtain an approximate decoherence time $t_d$, which is the time required to produce a certain amount $S_d$ (e.g., 1 bit for a Schr\"odinger Cat state) of entropy from an initial pure state:
\begin{equation}
t_d \simeq \frac{1}{\lambda}\left(S_d - \log{(\kappa\Delta x^2)} - \log{\theta}\right)
\label{Thumbnail}
\end{equation}
The key point is that $t_d$ is inversely linear in $\lambda$, but only logarithmic in the coupling $\theta$ or the initial properties of the environment (contained in $\kappa$).  As the coupling becomes very weak, then, we expect only moderate increases in the decoherence time.  This should be contrasted with QBM, where the decoherence time has a power-law dependence on the coupling strength, so that isolating the system from the environment (while still very difficult \cite{Zurek84ASI}) is not quite as hopeless as in the case studied here.

\begin{figure*}
\includegraphics[width=5in]{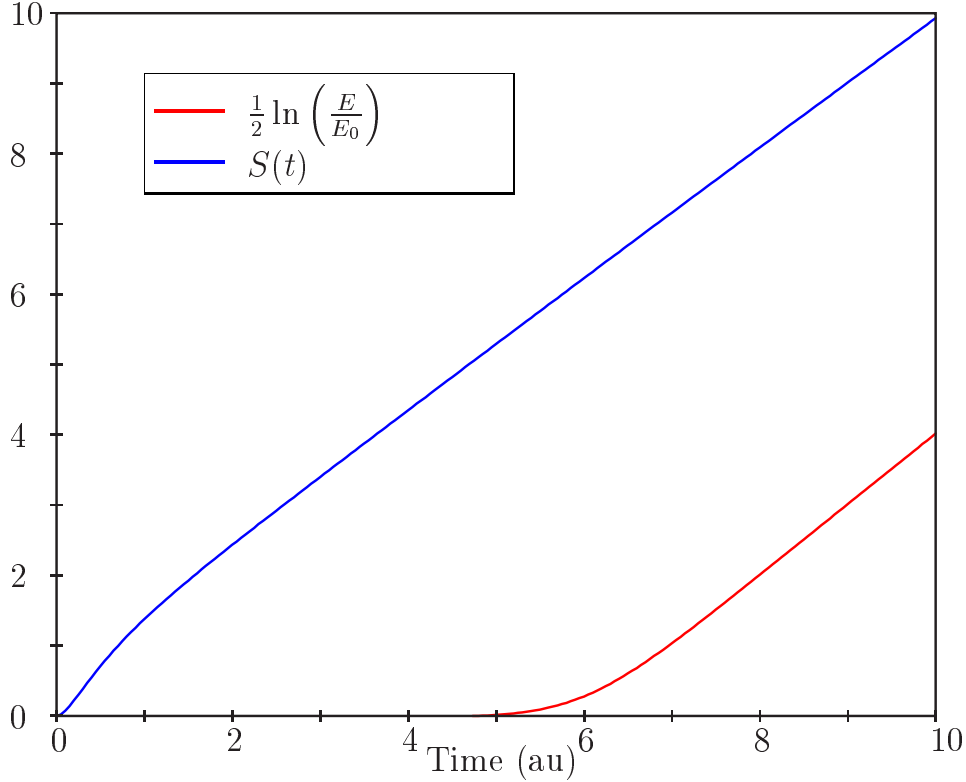}
\caption{{\em Comparison of $\entropy(t)$ and $\energy(t)$ for an extended state.}  Since $\energy$ grows as $e^{2\lambda t}$, we have plotted $\frac12\ln(\energy/\energy_0)$ so that both curves have the same slope asymptotically; thus all quantities plotted are dimensionless.  The noteworthy regime is $t<5$, where the entropy grows steadily but $\energy$ remains virtually constant.  The parameters for this calculation are: $\omega=10^{-5}$, $\lambda=1$, $m_s=m_e=1$, and $\theta=\frac{\pi}{512}$.  The initial state of the system is squeezed by a factor of $10^4$, with the long axis located at an angle of $\frac{\pi}{64}$ to the $\hat{x}$-axis; the environment is initially squeezed by a factor of $16$ in $\hat{p}$.}
\label{figSE}
\end{figure*}

Finally, we consider the system's energy.  If there is no coupling to the environment, then $\energy$ is a constant of the motion; conversely, when $\energy$ is substantially disrupted, the effects of the environment have clearly become noticeable at classical scales.  Since the system energy is given by
\begin{equation}
\energy(t) = \frac12\left(m_s\omega^2\expect{\hat{x}^2} + \frac{1}{m_s}\expect{\hat{p}^2}\right),
\end{equation}
we can immediately calculate its time derivative as
\begin{equation}
\pdiff{}{t}\energy(t) = \frac{f_1(t)-\gamma\expect{\hat{p}^2}}{m_s}
\end{equation}
Clearly $\energy$ will grow, because of the $f_1(t)$ term.  However, $\pdiff{A^2}{t}$ also contains $f_1$, but multiplied by $\Delta x^2$.  Thus, for states that are highly delocalized (Schr\"odinger cat states), rapid growth in $\entropy$ is achieved relative to the growth in $\energy$.  In addition, $\energy(0)$ (the initial energy of the system) plays no role in $\pdiff{\energy}{t}$; thus, if the initial energy of the system is high, the added energy $\energy(t)-\energy(0)$ will be negligible in comparison for some period of time.  We conclude that for initial states that have large $\Delta x^2$ and large $\energy(0)$ (i.e., superposition states over classical length scales), a relatively long period of time exists over which energy growth is negligible, while entropy grows rapidly.  An example is given in Figure \ref{figSE}.

\section{Conclusion}
We have studied an oversimplified model for a chaotic environment.  Our analysis is exact, although in the end we make simplifying assumptions in order to get ``thumbnail results'' (such as equation (\ref{Thumbnail})).  However, the conclusion that we are led to is significant, and we expect it to be quite generally applicable; we have shown that an unstable environment with a single degree of freedom can produce decoherence more readily than the canonical QBM environment, which requires infinitely many degrees of freedom and a much larger Hilbert space.  Our analysis necessarily examines a system that has an infinite-dimensional Hilbert space (since the phase space volume of the IHE is unbounded), but we can imagine a system with a bounded phase space that simulates an inverted oscillator arbitrarily well up to a certain time.  Such a system would produce the same results over the time of interest, but would not be analytically soluble beyond that time.

A brief comment is in order on the principle that enables our ``small'' (in terms of degrees of freedom and available Hilbert space dimension) environment to be so effective.  A good measure of an environment's decohering efficacy is the amount of entropy it can produce in the system, and the rate at which that entropy is produced.  For pure initial states of the environment, entropy can only be produced through entanglement; the total entropy that can be produced is limited by size of the environment's Hilbert space.  Thus, in the long run, larger environments can decohere more effectively.  The rate at which the entropy is produced, however, is limited by the rate at which the environment can explore its phase space.  Because a collection of harmonic oscillators is a stable system, small perturbations due to the state of the coupled system do not induce exploration of a large volume of phase space for any one oscillator.  The inverted oscillator, on the other hand, can explore its volume much more efficiently when perturbed.

We view this work as the first step in a larger project: understanding the role that chaotic systems play in decoherence.  These results for the IHE environment clearly show that a particular unstable environment can not only decohere a system in a manner markedly different from the standard (QBM) models of decoherence, but also display unexpected behavior -- e.g., the periodic breakdown of the  master equation formalism due to singularities in the coefficients.  While we believe we have sketched accurately the regime in which our toy model represents faithfully some aspects of the behavior of actual chaotic environments, the clear next step is to examine such environments directly using numerical simulations, and to see which of these results for the IHE remain unchanged.  Numerical studies carried out to date show a range of behaviors \cite{SakagamiKubotaniOkamura96ProgTP, KubotaniOkamuraSakagami95PhysicaA, MillerSarkar99PRE,MonteolivaPaz00PRL,MillerSarkarZarum98APPB,Zurek82PRD}.  We hope that the exactly solvable model we have described will aid in the analysis of the relevance of chaos to decoherence.  There remains another extremely important regime that we do not examine here: environments with many chaotic degrees of freedom (the atmosphere of the earth, for instance).

In addition to these obvious future directions, this work raises related questions about the behavior of open systems.  The key step in any decoherence process is that of tracing over the environment; if this is not done, the state of the supersystem is an entangled pure state, not a mixed state.  This is commonly justified by the argument that the environment is vast -- the whole universe, potentially -- and this size guarantees that some of the information that has been transferred to the environment has been irreversibly lost.  Despite this (reasonable) justification, explicit models of open quantum systems are usually bipartite; there is a small system, and a larger but still small (as compared to the rest of the universe) environment to which the system is coupled.  If, however, one imagines a set of ``concentric'' environments $\envsymbol^{(i)}$, each much larger than the last, as a model for the entire universe, then the environment to which the system is coupled ($\envsymbol^{(0)}$) serves not as an independent environment, but rather as a ``communication channel'' between the system and the greater universe (see \cite{Zurek82PRD,Zurek84ASI}, also \cite{Zurek03RMP}).  In this model, which we intend to investigate, the entropy of the system is not limited by the size of the environment (as it is in the bipartite system - environment) model, but only by its own size; a small local environment can lead to redundant records of the preferred observables of the system in the rest of the universe.  Our analysis of the IHE model indicates that chaotic local environments may act as ``amplifiers'' \cite{Glauber86} and, thus, carry information away from the system more efficiently than integrable environments.

\begin{acknowledgments}
The authors would like to thank Diego Dalvit, Juan Pablo Paz and Bill Unruh, for discussions and ideas.  This work was supported in part by a grant from the NSA.
\end{acknowledgments}

\bibliographystyle{apsrev}
\bibliography{decoherence}
\end{document}